\documentclass[twocolumn,showpacs,
amsmath,amssymb]{revtex4}
\usepackage{amssymb}
\usepackage[dvips]{graphicx}
\begin{document}

\title{ Theory of giant and nil proximity effects in cuprate semiconductors}
\author{A. S. Alexandrov}

\affiliation{Department of Physics, Loughborough University,
Loughborough LE11 3TU, United Kingdom\\}

\begin{abstract}
A  number of  observations point to the
possibility that high-$T_{c}$ cuprate superconductors may not be
conventional Bardeen-Cooper-Schrieffer (BCS) superconductors, but
rather derive from the Bose-Einstein condensation (BEC) of
real-space pairs, which are mobile small bipolarons. A solution of
the Gross-Pitaevskii equation describing  bose-condensate tunnelling
into a cuprate semiconductor is analytically found. It accounts
qualitatively and quantitatively for  nil and giant proximity
effects  discovered experimentally in cuprates.

\end{abstract}


\maketitle

Perhaps the most striking property of cuprate superconductors is the
Giant Proximity Effect (GPE). Several groups \cite{bozp} reported
that in the Josephson cuprate $SNS$ junctions supercurrent can run
through normal $N$-barriers with the thickness $2L$ greatly
exceeding the coherence length, when the barrier is made from a
slightly doped non-superconducting cuprate (the so-called $N'$
barrier). Using an advanced molecular beam epitaxy, Bozovic \emph{et
al.} \cite{bozp} proved that GPE is intrinsic, rather than  caused
by any extrinsic inhomogeneity of the barrier. Resonant scattering
of soft-x-ray radiation did not find any signs of intrinsic
inhomogeneity (such as charge stripes, charge-density waves, etc.)
either \cite{boz5}. Hence GPE defies the conventional explanation,
which predicts that the critical current should exponentially decay
with the characteristic length of about the coherence length, $\xi
\lesssim 1$ nm in  cuprates. Annealing the junctions at low
temperatures in vacuum rendered the barrier insulating. Remarkably
when  the $SN'S$ junction was converted into a
superconductor-insulator-superconductor (SIS) device no supercurrent
was observed, even in devices with the thinnest (one unit cell
thick) barriers \cite{boz0} (nil proximity effect, NPE).

 Cuprate superconductors  with low
density of free carriers and poor mobility (at least in one
direction) are characterized by poor screening of high-frequency
c-axis polarised optical phonons. The unscreened Fr\"ohlich
interaction between oxygen holes and these phonons combined with
on-site repulsive correlations (Hubbard $U$)  binds holes into
superlight intersite bipolarons \cite{ale5}, which are real-space
pairs dressed by phonons. Experimental evidence for  exceptionally
strong e-ph interactions
 is now so overwhelming \cite{mic1,ita,tal,zhao,LAN} that a bipolaronic charged bose gas
 (CBG) \cite{alebook1}
 could be a feasible alternative to  the BCS-like
scenarios of cuprates. The bipolaron theory predicted such key
features of cuprate superconductors as anomalous upper critical
fields, spin and charge pseudogaps, and unusual isotope effects
later discovered experimentally. The theory explained normal state
kinetics, including the anomalous Hall-Lorenz number, high $T_c$
values, specific heat anomalies of cuprates (for a  review see
\cite{alebook1}), and more recently the d-wave symmetry of the order
parameter in underdoped \cite{alesym} and overdoped \cite{andsym}
samples, the normal state Nernst effect \cite{alezav} and
diamagnetism \cite{aledia}.

Here I show that both GPE and NPE can be broadly understood as the
BEC tunnelling into a cuprate \emph{semiconductor}.

  A stationary condensate wave function of CBG, $\psi(\bf r)$,
 obeys the following  equation,
\begin{equation}
\left[-{\Delta\over{2m}} - \mu + V({\bf r})\right]\psi({\bf r}) =0,
\end{equation}
where $m$ is the boson mass, $\mu$ is  the chemical potential and
$V({\bf r})$ describes
 the Coulomb and hard-core composed boson-boson repulsions, and their attraction to a neutralising charge background
  (here and further
$\hbar=c=k_B=1$). While the electric field potential can be found
from the corresponding Poisson-like equation \cite{alevor}, a
solution of two coupled nonlinear differential equations for the
order parameter $\psi({\bf r})$ and for the potential $V({\bf r})$
is a nontrivial mathematical problem, which requires a
"multishooting" numerical approach. Here we restrict our analysis by
a short-range potential, $V({\bf r})=V |\psi({\bf r})|^2$, where a
constant $V$ accounts for the hard-core and screened Coulomb
repulsions. Then Eq.(1) is the familiar Gross-Pitevskii (GP)
equation \cite{gro}. In the tunnelling geometry of $SN'S$ junctions,
Figs.1,2, it takes the form,
\begin{equation}
{1\over{2m_c}}{d^2\psi(Z)\over{dZ^2}}=[V |\psi(Z)|^2-\mu]\psi(Z),
\end{equation}
in the superconducting region, $Z<0$, near the $SN$ boundary, Fig.1.
Here  $m_c$ is the boson mass in the direction of tunnelling along
$Z$. Deep inside the superconductor $|\psi(Z)|^2=n_s$ and $\mu=Vn_s$
, where the condensate density $n_s$ is about $x/2$, if the
temperature is well below $T_c$  of the superconducting electrode.
 Here the in-plane lattice constant $a$ and the unit cell volume are
 taken as unity, and $x$ is the doping level as in La$_{2-x}$Sr$_{x}$CuO$_4$.

\begin{figure}
\begin{center}
\includegraphics[angle=-90,width=0.55\textwidth]{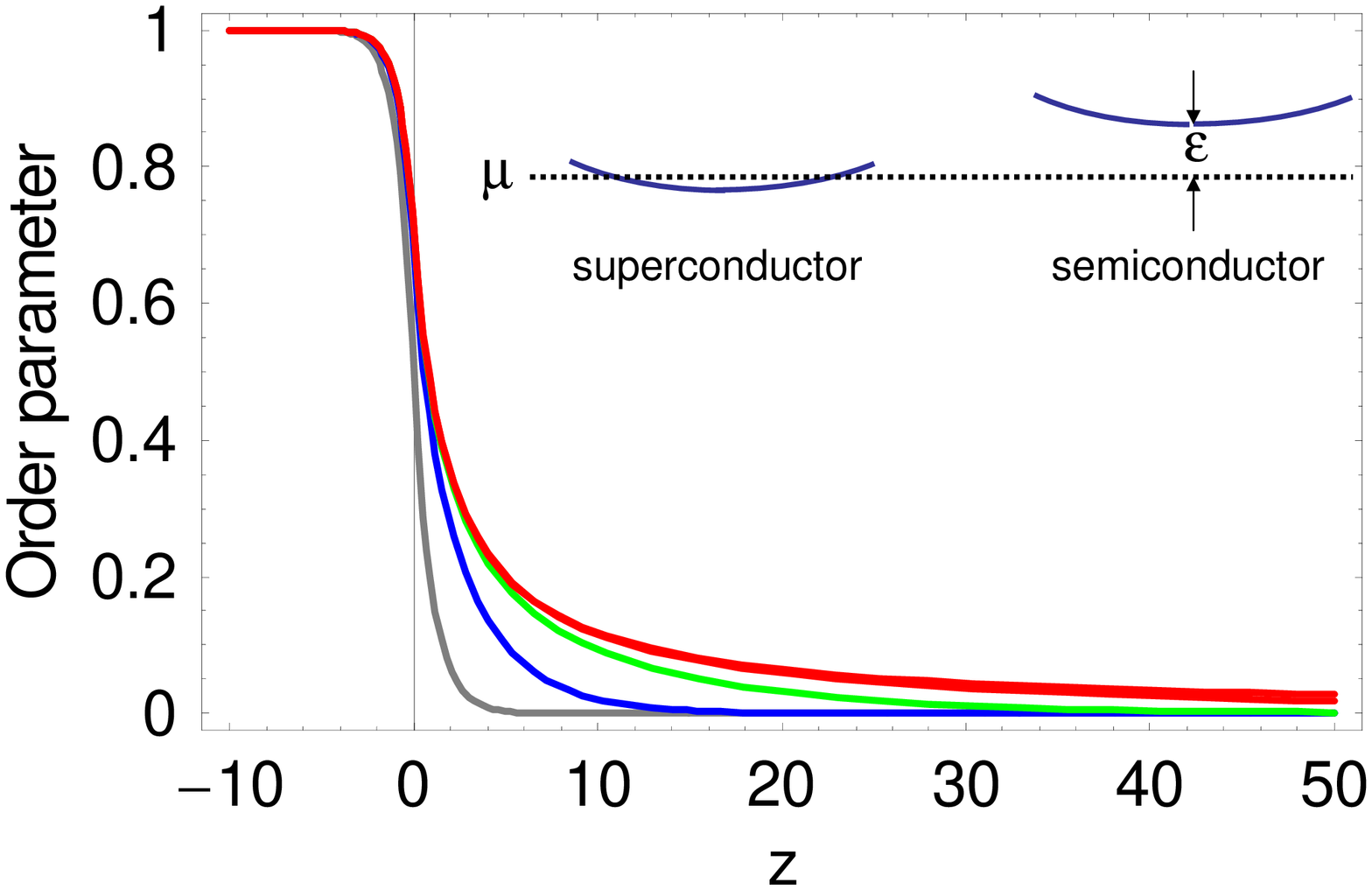}
\vskip -0.5mm \caption{BEC order parameter at the $SN$ boundary for
$\tilde{\mu}=1.0,0.1,0.01$ and $ \leqslant 0.001$ (upper curve). The
chemical potential is found above the boson band-edge due to the
boson-boson repulsion in  cuprate superconductors and below the edge
in   cuprate semiconductors with low doping.}
\end{center}
\end{figure}

The normal barrier  at $Z
>0$ is an underdoped cuprate above its
transition temperature, $T'_c<T$ where the chemical potential $\mu$
lies below the bosonic band by some energy $\epsilon$, Fig.1, found
from
\begin{equation}
\int dE N(E)[\exp((E+\epsilon)/T)-1]^{-1}=x'/2.
\end{equation}
Here $N(E)$ is the bipolaron density of states (DOS), and $x' < x$
is the doping level of the barrier. In-plane bipolarons are
quasi-two dimensional bosons, propagating along the $CuO_2$ planes
with the effective mass $m$ several orders of magnitude smaller than
their out-of-plane mass, $m_c\gg m$, which accounts for the
remarkable anysotropy of cuprates at low temperatures
\cite{alemotkab}. Using bipolaron band dispersion, $E({\bf
K})=K^{2}/2m+2t_{c}[1-\cos(K_{\perp}d)]$,  the density of states is
found as  $N(E)=(m/2\pi^2)\arccos(1-E/2t_c)$ for $0<E<4t_{c}$, and
$N(E)=m/2\pi$
 for $4t_c<E$. Here $K$ and $K_\perp$ are the in-plane and out-of-plane center-of-mass momenta, respectively, $t_c= 1/2m_cd^2$, and  $d$ is the inter-plane
 distance. Assuming that the critical temperature of the barrier $T'_c \approx T_0/\ln(T_0/2t_c)$ (here   $T_0=\pi x'/m \gg T'_c$) is much larger than $t_c$
one obtains from Eq.(3)
\begin{equation}
\epsilon(T)\leqslant -T\ln(1-e^{-T_0/T}),
\end{equation}
which turns into zero at $T=T'_c$.

 Then the GP equation in the barrier is
written as
\begin{equation}
{1\over{2m_c}}{d^2\psi(Z)\over{dZ^2}} =[V
|\psi(Z)|^2+\epsilon]\psi(Z).
\end{equation}
Introducing the bulk coherence length, $\xi= 1/(2m_c n_sV)^{1/2}$
and dimensionless $f(z)=\psi(Z)/n_s^{1/2}$,
$\tilde{\mu}=\epsilon/n_sV$, and $z=Z/\xi$ one obtains
 for a real
$f(z)$
\begin{equation}
{d^2f\over{dz^2}} =f^3-f,
\end{equation}
if $z<0$, and
\begin{equation}
{d^2f\over{dz^2}}=f^3+\tilde{\mu}f,
\end{equation}
if
 $z>0$. These equations can be readily solved using  first
integrals of motion respecting the boundary conditions,
$f(-\infty)=1$, and $f(\infty)=0$,
\begin{equation}
{df\over{dz}}= -(1/2+f^4/2-f^2)^{1/2},
\end{equation}
and
\begin{equation}
{df\over{dz}}= -(\tilde{\mu}f^2+f^4/2)^{1/2},
\end{equation}
for $z<0$ and $z>0$, respectively. The solution in the
superconducting electrode is given by
\begin{equation}
f(z)=\tanh \left[-2^{1/2}z+0.5
\ln{{2^{1/2}(1+\tilde{\mu})^{1/2}+1}\over{2^{1/2}(1+\tilde{\mu})^{1/2}-1}}\right].
\end{equation}
It decays  in the close vicinity of the barrier from 1 to
$f(0)=[2(1+\tilde{\mu})]^{-1/2}$ in the interval about the coherence
length $\xi$. On the other side  of the boundary, $z>0$, it is given
by
\begin{equation}
f(z)={(2\tilde{\mu})^{1/2}\over{\sinh\{z\tilde{\mu}^{1/2}+\ln[2(\tilde{\mu}(1+\tilde{\mu}))^{1/2}+(1+4\tilde{\mu}(1+\tilde{\mu}))^{1/2}]\}}}
.
\end{equation}
Its profile is shown in Fig.1. Remarkably, the order parameter
penetrates  the normal layer up to the length $Z^* \thickapprox
(\tilde{\mu})^{-1/2}\xi$, which could be larger than $\xi$ by many
orders of magnitude,  if $\tilde{\mu}$ is  small. It is indeed the
case, if the barrier layer is sufficiently doped. For example,
taking $x'=0.1$,   c-axis $m_c=2000 m_e$, in-plane $m=10 m_e$
\cite{alebook1}, $a=0.4$ nm, and $\xi=0.6$ nm, yields
 $T_0\approx 140$ K and $(\tilde{\mu})^{-1/2}\gtrsim 50$ at $T=25$K. Hence the
 order parameter could penetrate  the normal cuprate semiconductor
 up to  a hundred coherence lengths or even more as observed \cite{bozp}. If the thickness of the barrier $L$ is small compared with $Z^*$,
and $(\tilde{\mu})^{1/2}\ll 1$, the order parameter decays following
 the power law, rather than exponentially,
\begin{equation}
f(z)={\sqrt{2}\over{z+2}}.
\end{equation}
Hence, for $L \lesssim Z^*$, the critical current should also decay
following the power law as discussed below. On the other hand, for
an \emph{undoped}
 barrier $\tilde{\mu}$ becomes
 larger than unity, $\tilde{\mu}\varpropto \ln(mT/\pi x')\rightarrow \infty$ for any finite temperature $T$  when $x' \rightarrow
 0$, and the current should exponentially decay with the characteristic length  smaller that $\xi$, as is experimentally observed as well \cite{boz0}.

To get more insight in the temperature and barrier-thickness
dependence of the critical current  one has to solve the GP equation
in the $SN'S$ junction geometry, Fig.2, with the current,
$J=(2en_s/m_c\xi)R^2 d\Theta/dz$. Here  $R(z)$ and $\Theta(z)$ are
the amplitude and the phase of the order parameter, respectively,
$f(z)=R(z)\exp(i\Theta(z))$. The current does not depend on $z$, so
that the equation can be written as
\begin{equation}
{d^2R\over{dz^2}} =R^3-(1-j^2/R^4)R,
\end{equation}
if $|z|>l$, and
\begin{equation}
{d^2R\over{dz^2}}=R^3+(\tilde{\mu}+j^2/R^4)R,
\end{equation}
if $|z|<l$, where $l=L/\xi$ and $j=Jm_c\xi/2en_s$. Using  integrals
of motion respecting the boundary conditions,  $dR/dz=0$ at $z=0$ (
$R(z)=R(-z)$ since the junction is symmetric),
$R_{\infty}^3=R_{\infty}-j^2/R_{\infty}^3$, and $dR_\infty/dz=0$,
these equations are reduced to
\begin{equation}
{dR\over dz}=[(R^4-R_0^4)/2+\tilde{\mu}
(R^2-R_0^2)+j^2(R_0^{-2}-R^{-2})]^{1/2},
\end{equation}
for $0<z<l$, and
\begin{equation}
{dR\over dz}=[(R^4-R_{\infty}^4)/2+
(R_{\infty}^2-R^2)+j^2(R_{\infty}^{-2}-R^{-2})]^{1/2},
\end{equation}
for $z>l$.  Then integrating Eq.(14) from $z=0$ up to $z=l$ allows
us to connect the current and the order parameter, $R_0$ in the
center of the barrier at $z=0$,
\begin{equation}
\int_{1}^{R_l^2/R_0^2}{dx\over{\sqrt{x(x^2-1)+x(x-1)\alpha+(x-1)\beta}}}
=\sqrt{2}R_0l.
\end{equation}

\begin{figure}
\begin{center}
\includegraphics[angle=-90,width=0.50\textwidth]{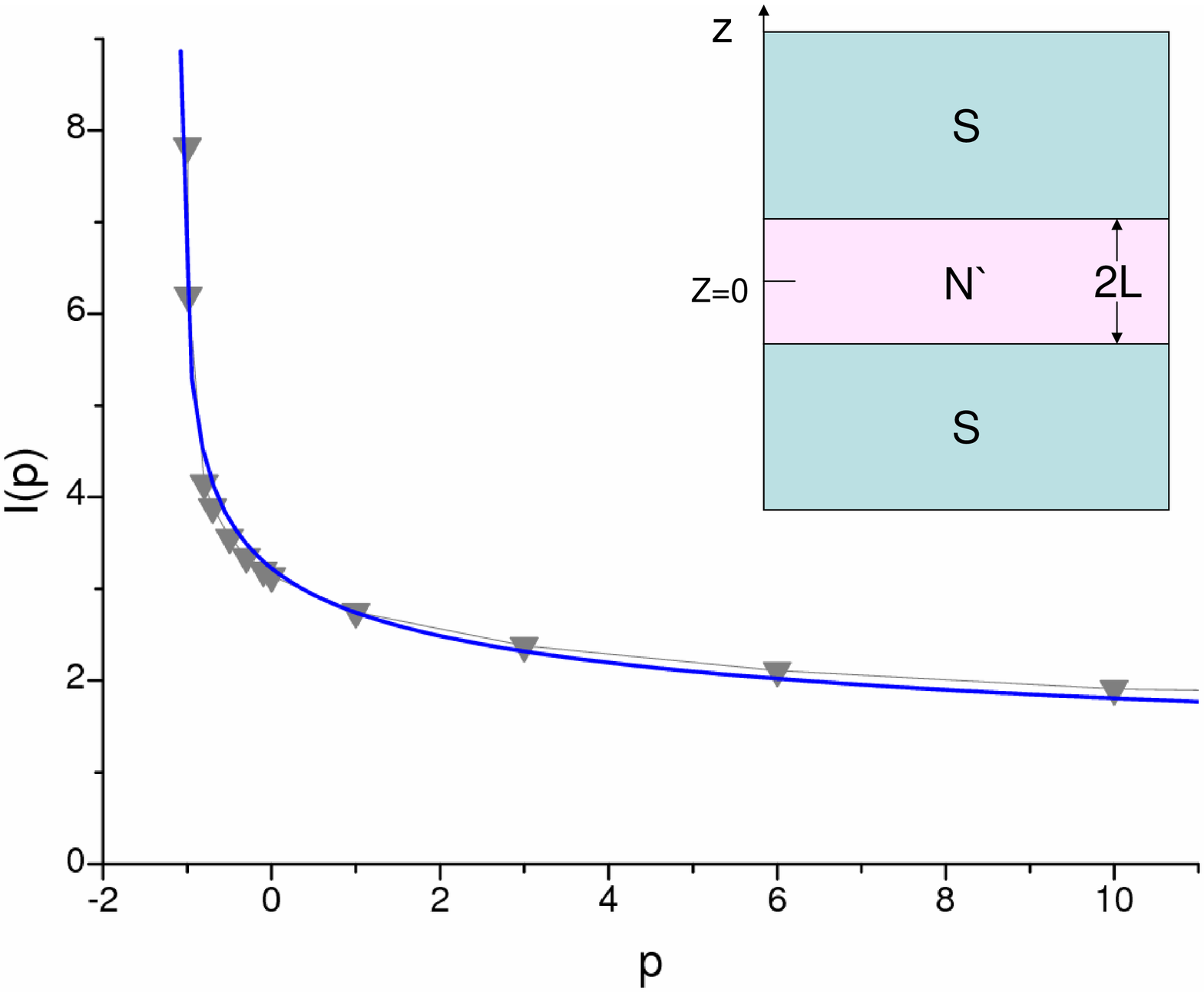}
\vskip -0.5mm \caption{Numerical values of the intgral $I(p)$ in
Eq.(19) (symbols) compared with the analytical approximation (solid
line). Inset shows schematically the trilayer  $SN'S$ device of
Ref.\cite{bozp}.}
\end{center}
\end{figure}

Since $R_l\gg R_0$ for sufficiently thick barrier, $l\gg 1$,  the
upper limit in this integral can be extended to $\infty$ yielding a
transcendental equation for $R_0$ as a function of $j$
\begin{eqnarray}
&& _2F_1
\left[1/2,1/2;1;{1\over{2}}(1-{3+\alpha\over{2\sqrt{\alpha+\beta
+2}}})\right]\cr
&=&\sqrt{2}(\alpha+ \beta +2)^{1/4}R_0l/\pi,
\end{eqnarray}
where $_2F_1(a,b;c;z)$ is the Gauss hypergeometric function,
$\alpha=2\tilde{\mu}/R_0^2$ and $\beta=2j^2/R_0^6 $. When
$\alpha=\beta=0$, one obtains $R_0 \propto 1/l$, as in Eq.(12), and
when $\alpha \rightarrow \infty$, the order parameter decays
exponentially as in Eq.(11).

To evaluate the critical current one can reduce the integral in
Eq.(17) to a single parameter integral,  $I(p)= \int_0^{\infty}
dx[x(x+1)^2+px]^{-1/2}$ with $p=-1+4(\alpha+\beta+2)/(3+\alpha)^2$.
This integral can be analytically approximated (see Fig.2) as  $
I(p) \approx 3.3(1.1+p)^{-1/4}$, if $p$ is not too close to $-1$.
Then solving
\begin{equation}
 I(p) =R_0l(3+\alpha)^{1/2},
 \end{equation}
   with
respect to the current yields $j \approx R_0 (25/2l^4
-R_0^4-\tilde{\mu}R_0^2)^{1/2}$, and the critical current
$j_c\approx 4(1-2\tilde{\mu}l^2)/l^3$. These expressions are applied
when $l\lesssim (\tilde{\mu})^{-1/2}$ (i.e. $L \lesssim Z^*$).

To get the temperature dependence of the critical current $J_c (T)$
in the whole range of parameters one can apply the scaling, $j_c
\propto R_0^2/l$, where $R_0$ is defined as the exact solution of
the $SN'$ boundary problem, Eq.(10) at $z=l$,  $R_0=f(l)$. In such a
way one finds  (in ordinary units)
\begin{equation}
J_c(T) = {A\epsilon (T)\over{k_B\sinh^2[L\sqrt{2m_c\epsilon
(T)}/\hbar]}},
\end{equation}
where $A \approx 4ek_B\xi_0^2 n_{s0}/\hbar L$ is temperature
independent. With the zero-temperature coherence length $\xi_0 =
0.5$ nm and condensate density $n_{s0}=x/2\Omega$ one estimates $A
\approx 150$ kA/cm$^2$K for optimum doping $x=0.15$ and the barrier
thicness $2L=10$ nm (here $\Omega$ is the unit cell volume). Eq.(20)
fits well the temperature dependence and absolute values of the
critical current, Fig.3, measured by Bozovic et al. \cite{bozp} with
$T_0=160$ K, and $m_c=1000 m_e$, Fig.3.
\begin{figure}
\begin{center}
\includegraphics[angle=-90,width=0.55\textwidth]{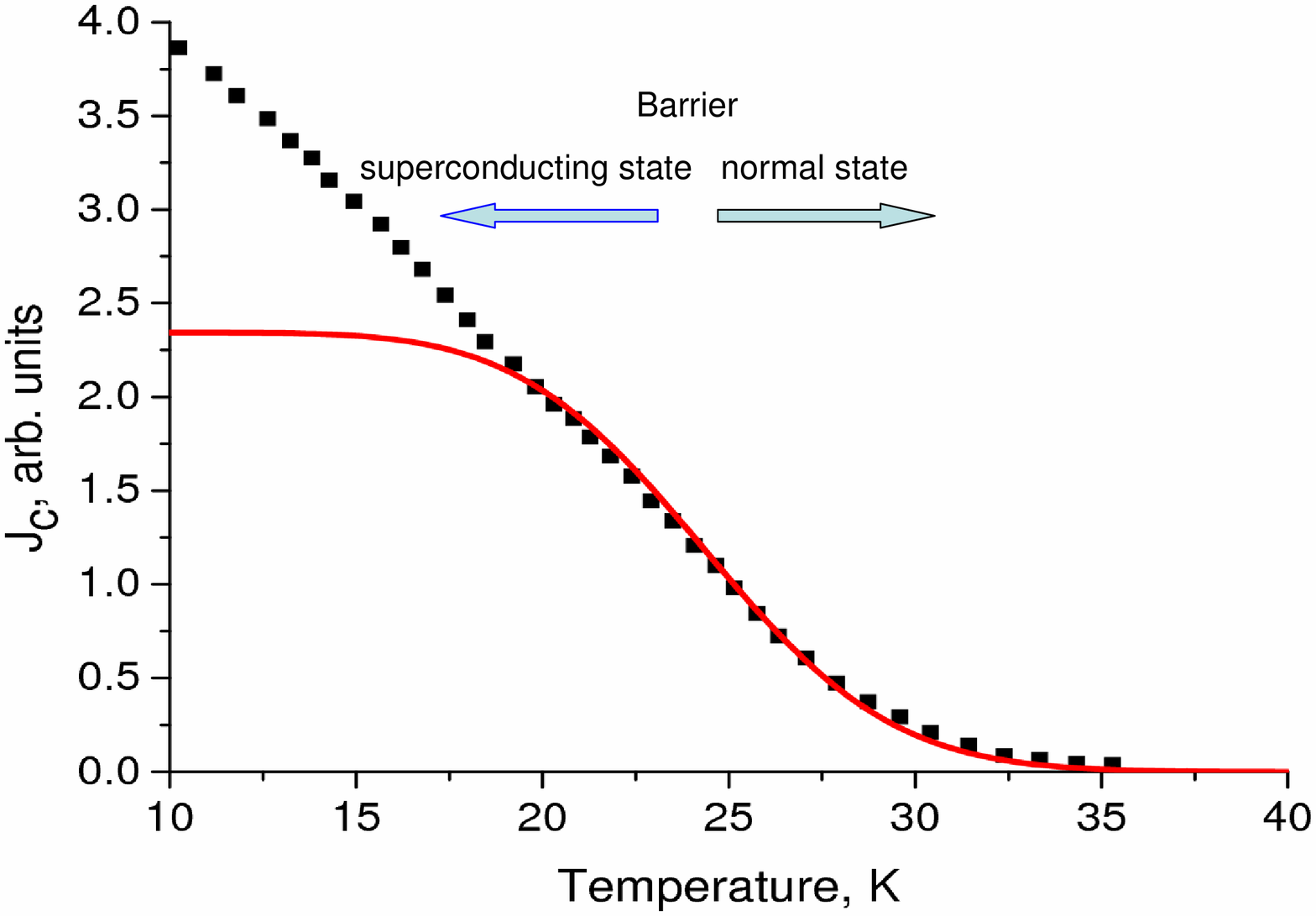}
\vskip -0.5mm \caption{Temperature dependence of the critical
current of the trilayer $SN'S$ device \cite{bozp} with $L=5$ nm
(symbols) described by Eq.(20) at $T> T_c'$.}
\end{center}
\end{figure}
The typical device of Ref.\cite{bozp}  used
La$_{1.85}$Sr$_{0.15}$CuO$_4$ with $T_c \approx 45 $K as the
superconducting electrodes while  the normal $N'$ barrier was made
of underdoped La$_{2}$CuO$_{4+\delta}$ with typical $T_c' \approx
25$K. One can see from Fig.3 that the theory describes the critical
current in the  normal region of the barrier, where $\tilde{\mu}$ is
positive. When the barrier becomes superconducting (c.a. below $20$
K) the experimental critical current naturally exceeds the
theoretical "semiconducting" $J_c(T)$.

A possibility of real-space pairing, proposed originally by Ogg
\cite{ogg} and later on by Schafroth  and Blatt and Butler
\cite{but}   has been the subject of many discussions as opposed to
the Cooper pairing, particularly heated over the last 20 years after
the discovery of high temperature superconductivity in cuprates. Our
extension of the BCS theory towards the strong interaction between
electrons and ion vibrations proved that BCS and Ogg-Schafroth
pictures are two extreme limits of the same problem. For a very
strong electron-phonon coupling, polarons become self-trapped on a
single lattice site and bipolarons are on-site singlets. In the
Holstein model of the electron-phonon interaction
 their mass  appears only in the second order of polaron hopping,
 so that on-site bipolarons are very heavy. This estimate led some authors to
 the conclusion that the formation
of itinerant small polarons and bipolarons in real materials is
unlikely \cite{mel}, and high-temperature bipolaronic
superconductivity is impossible \cite{and2}. However, we have
noticed that the Holstein model is an extreme polaron model, and
typically yields the highest possible value of the (bi)polaron mass
in the strong coupling regime.  Cuprates are characterized by poor
screening of high-frequency optical phonons and are more
appropriately described by the long-range Fr\"ohlich electron-phonon
interaction \cite{ale5}. The unscreened Fr\"ohlich electron-phonon
interaction provides relatively light small polarons and bipolarons,
which are several orders of magnitude lighter than small Holstein
(bi)polarons.

I conclude that the bipolaron theory accounts for GPE and NPE
 in slightly doped semiconducting and undoped insulating
cuprates, respectively. It predicts  the occurrence of a new length
scale, $\hbar/\sqrt{2m_c\epsilon (T)}$.  In a wide temperature range
 far from the transition point, $T'_c <T < T_0$,  this length
turns out much larger than the zero-temperature coherence length, if
bosons are almost two-dimensional (2D). The physical reason, why the
quasi-2D bosons display a large normal-state coherence length,
whereas 3D Bose-systems (or any-D Fermi-systems) at the same values
of parameters do not, originates in the large DOS near the band edge
of two-dimensional bosons. Since  DOS is large, the chemical
potential is pinned near the edge with the magnitude, $\epsilon
(T)$, which is  exponentially small when $T<T_0$. Importantly the
theory predicts an unusual dependence of $J_c$ on the barrier
thickness $2L$, in particular at low temperatures, $T<T_c'$, when
the barrier is superconducting, there is almost no $L$ dependence,
above the transition of the barrier, $J_c \propto 1/L^3$, and well
above the transition ($T\gtrsim T_0
>> T_c'$) $J_c$  decays exponentially with $L$. Based on a great
number of experimental observations \cite{alebook1} including GPE
and NPE I argue that the most likely scenario for superconducting
cuprates is the genuin
 Bose-Einstein condensation  of real-space mobile
 lattice bipolarons.

  I thank  A.F. Andreev, I. Bozovic, L.P.
Gor'kov,  V.V. Kabanov for valuable discussions, and E.V.
Ferapontov, S. Flach, V. Khodel, F.V. Kusmartsev, and A.P. Veselov
for illuminating comments on integrability conditions for
differential systems. The work was supported by EPSRC (UK) (grants
no. EP/C518365/1, EP/D035589/1, and EP/D07777X/1
 ).

\end{document}